\documentstyle[11pt]{article}
\begin{document}
\renewcommand{\theequation}{\thesection.\arabic{equation}}
\renewcommand{\refname}{References.}
\newcommand{\sect}[1]{ \section{#1} \setcounter{equation}{0} }
\title{The $\beta$-function of the chiral Gross Neveu model at $O(1/N^2)$.}
\author{J.A. Gracey, \\ Department of Applied Mathematics and Theoretical
Physics, \\ University of Liverpool, \\ P.O. Box 147, \\ Liverpool, \\
L69 3BX, \\ United Kingdom.}
\date{}
\maketitle
\vspace{5cm}
\noindent
{\bf Abstract.} We compute the $O(1/N^2)$ correction to the critical exponent
$2\lambda$ $=$ $-$ $\beta^\prime(g_c)$ for the chiral Gross Neveu model in
arbitrary dimensions by substituting the corrections to the asymptotic scaling
forms of the propagators into the Schwinger Dyson equations and solving the
resulting consistency equations.

\vspace{-16cm}
\hspace{10cm}
{\bf LTH-335}
\newpage
\sect{Introduction.}
The large $N$ expansion has proved to be a useful technique in providing an
alternative method of analysing renormalizable quantum field theories which
possess $N$ fundamental fields. For example, it was demonstrated in \cite{1}
that the (perturbatively massless) two dimensional Gross Neveu models with
discrete and continuous chiral symmetry exhibit dynamical mass generation as
well as being asymptotically free field theories with this technique. This
former property cannot be accessed in conventional perturbation theory and such
toy models have proved to be important as a step to understanding similar
phenomena in complicated four dimensional gauge theories such as quantum
chromodynamics. One of the limitations, however, of a conventional large $N$
analysis is that one cannot proceed much beyond the leading order due in part
to the appearance of intractable integrals whose treatment is hindered
primarily by the presence of the dynamically generated mass.

Methods to overcome this problem have been developed in \cite{2,3} and applied
initially to the $O(N)$ bosonic $\sigma$ model to solve that theory to
$O(1/N^2)$. The technique exploits the properties of the field theory at its
$d$-dimensional critical point which is defined to be the non-trivial zero
of the $\beta$-function. The simplifying feature of studying a model there is
the additional symmetry present. As the $\beta$-function vanishes the Green's
function of the fields are conformally invariant and obey simple power law
behaviour which, of course, has a massless form. In essence one is using the
conformal bootstrap approach to solve the field theory, \cite{C}. As a
consequence, one avoids one of the difficulties of the conventional large $N$
approach in that the previously intractable integrals become calculable using
techniques developed to compute massless Feynman diagrams, \cite{4}. This
technique, known as uniqueness, is in effect a method of conformal integration
in $d$-dimensions.  Essentially when the sum of the exponents of bosonic
propagators at a $3$-vertex in coordinate space sum to the dimension of space
time, the vertex is replaced by a triangle of propagators whose exponents are
related to the original vertex. This represents an integration within the
overall Feynman diagram. In essence it is the Yang Baxter star triangle
relation, but is more general since it is valid in arbitrary (fixed) space time
dimension. Moreover, as one is now dealing with a scale invariant situation the
problem of solving the model becomes one of computing the critical indices or
exponents of the fields which will have an anomalous portion in addition to the
canonical dimension. The anomalous piece is the part which carries the quantum
properties of the model and from the universality criterion of statistical
mechanics (see, for example, \cite{5}), it will be a function of the spacetime
dimension and the basic parameters of the model, such as $N$ in the case of a
large $N$ expandable theory, and can be calculated order by order in $1/N$. The
equations which one uses to do this are deduced by representing the Dyson
equations of the Green's functions in the critical region of the theory from
which one can deduce self consistent equations whose solution yields the
exponents, \cite{2}.  The analytic expressions which result for the anomalous
dimensions can be related to the respective renormalization group functions
precisely at the critical point, \cite{2,5}. One important function which
characterizes a field theory is the $\beta$-function which governs how the
coupling constant behaves with the renormalization scale. Ordinarily one
calculates it order by order in powers of the perturbative coupling constant,
which is assumed to be small. In the large $N$ critical point approach which we
concentrate on here it can, however, be determined by computing the critical
exponent $2\lambda$ $=$ $-$ $\beta^\prime(g_c)$ where $g_c$ is the critical
point. Whilst this may seem to be an indirect way to proceed it is important to
note that the $O(1/N^2)$ corrections to the $\beta$-function of the $O(N)$
bosonic $\sigma$ model could only be deduced in this way, \cite{3}.

Following the pioneering work of \cite{2,3} the method has been extended to
analyse models with fermions, [7-11]. In particular the $O(N)$ Gross Neveu
model has been solved at $O(1/N^2)$ with the fermion and auxiliary field
anomalous dimensions, \cite{6,7}, and the $\beta$-function exponent $\lambda$,
\cite{8,9}, all having been computed in arbitrary dimensions. Now that matter
fields have been successfully incorporated within the critical point approach
it is possible to examine other models to the same precision. One such model
which is currently relevant and related to the $O(N)$ Gross Neveu model is the
chiral Gross Neveu model, \cite{1,11}, which possesses a continuous chiral
symmetry. This was first discussed in \cite{11} by Nambu and Jona-Lasinio in
four dimensions and later in \cite{12} where the connection with hadronic
physics was first introduced. Moreover, four fermi interactions have recently
been a subject of intense study as it provides a possible alternative to the
Higgs mechanism in the standard model. There a fermion fermion bound state
plays the role of the Higgs boson, \cite{13,14} and so it is important to have
as complete a picture as possible of the quantum structure of the model.
Therefore, the aim of this paper is to compute the exponent $\lambda$ of the
chiral Gross Neveu or Nambu--Jona-Lasinio model at $O(1/N^2)$ which relates to
the $\beta$-function. This is one order beyond what has been computed before in
this model. One property that the chiral Gross Neveu model shares with other
models in two dimensions is that it possesses an exact $S$-matrix, \cite{A},
from which an expression for the exact mass gap has recently been deduced,
\cite{B}. The $S$-matrix has been constructed on the assumption that
$3$-particle $S$-matrix elements factorize into the product of the constituent
$2$-particle elements which is a consequence of the underlying Yang Baxter
triangle relation associated with integrable field theories. Whilst the model
is asymptotically free and therefore possesses the property of dimensional
transmutation in two dimensions, it might appear that one has already
determined the $\beta$-function of the $2$-dimensional model. However, such a
statement overlooks the implicit nature of the situation. As far as we are
aware, the explicit form of the $2$-dimensional $\beta$-function which can in
principle be deduced from the $S$-matrix spectrum, has never been achieved in
detail. In providing {\em new} information on the $O(1/N^2)$ structure of the
$\beta$-function here we are putting in place a means of checking any future
attempt at that difficult programme. However, more importantly, our motivation
in the present calculation goes beyond the two dimensional model for two
reasons. First, we are determining the $d$-dimensional structure of the
$\beta$-function by exploiting the conformal nature of the fixed point and
applying the $d$-dimensional generalization of the star-triangle relation, we
are able to deduce new information on the three dimensional model {\em
simultaneously}. Our result goes beyond anything which has been calculated
before in this area since we are computing at $O(1/N^2)$ and this is relevant
for recent lattice simulations of the three dimensional four fermi interactions
for relatively low values of $N$ which have been a topic of interest lately,
\cite{22}.

Further, the anomalous dimensions of all the fields of the theory have been
deduced in \cite{15} and by calculating $\lambda$ we will in fact solve the
model completely to second order since the remaining critical exponents
characterizing the model can be deduced through the hyperscaling relations
which have been checked at $O(1/N)$ in \cite{16} although their consistency
merely reflects the renormalizability of the model. Indeed the work we present
builds very much on the calculation of $\lambda$ in the $O(N)$ case, \cite{8}.
Although there are substantially more Feynman graphs to consider due to the
presence of another field it is possible to make use of results obtained for
those graphs in the continuous chirally symmetric case to minimize the work
needed to obtain new results. This is significant when one realises the basic
intractability that a conventional large $N$ analysis would entail. The
previous remarks are important in stating our second motivation on the need to
have $d$-dimensional results. One feature of four fermi theories with various
chiral symmetries is that they are believed to lie in the same universality
class as the Yukawa models with the same chiral structure. Such equivalences,
for example, are established using the technique of the $\epsilon$-expansion.
In essence the renormalization group functions, calculated to several orders in
perturbation theory, are expanded in powers of $\epsilon$, where $d$ $=$ $2$
$+$ $\epsilon$ or $4$ $-$ $2\epsilon$ depending upon which dimension the
coupling constant of the model is dimensionless in, and compared with the
analogous function in a similar expansion in the other model. As the critical
exponents calculated within the large $N$ self consistency approach discussed
here encode {\em all} orders information on perturbation theory, such
$d$-dimensional results have been important in confirming such equivalences in
the $O(N)$ case, \cite{22}.

The paper is organized as follows. In section 2, we review the basic formalism
we need for solving the chiral Gross Neveu model and recall the relevant
features of the $O(1/N)$ calculation of $\lambda$, which unlike the bosonic
$O(N)$ $\sigma$ model calculation at the same order requires the inclusion of
several $2$-loop Feynman graphs. We devote section 3 to the derivation of the
master equation whose solution will yield $\lambda$ at $O(1/N^2)$. This
involves considering several three and four loop graphs in the Dyson equations
of two of the fields of the model. The equation is solved in section 4 to
obtain an arbitrary dimensional expression for the exponent from which we
deduce information on the $3$-dimensional model as a simple corollary.

\sect{Preliminaries.}

The lagrangian of the model we will analyse is, \cite{1,11},
\begin{equation}
L ~=~ i \bar{\psi}^i {\partial \!\!\! /} \psi^i + \sigma \bar{\psi}^i \psi^i
+ i \pi \bar{\psi}^i \gamma^5\psi^i - \frac{1}{2g^2}(\sigma^2+\pi^2)
\end{equation}
where $\sigma$ and $\pi$ are auxiliary bosonic fields which when eliminated
from (2.1) yields a model with a four fermi interaction which has a $U(1)$
$\times$ $U(1)$ global chiral symmetry. The fermion field $\psi^i$ lies in an
$N$-tuplet, $1$ $\leq$ $i$ $\leq$ $N$, and the parameter $1/N$ will play the
role of the coupling constant in our analysis and we use the convention that
$\mbox{tr} \, 1$ $=$ $2$. In the usual approach to solving (2.1) in the large
$N$ expansion, \cite{1}, one can perform the path integral over the fermion
fields in (2.1) as they appear quadratically in the action. To proceed further
one uses the saddle point approximation when $N$ is large to deduce from the
effective potential that the true vacuum of the theory is not the perturbative
one but that where the $\sigma$ and $\pi$ fields become dynamical. Further, in
this vacuum a mass is generated for the fermion field although perturbatively
it remains massless, \cite{1}. Equipped with this structure one normally
proceeds to solve the model in the large $N$ expansion by renormalizing the
various Green's functions. Consequently, one can deduce the large $N$
approximation of the $\beta$-function, for example. To go beyond the leading
order turns out to be impossible due to the appearance of intractable integrals
whose divergence structure, handled by some regularization, cannot be
extracted. This is chiefly a consequence of the mass present in the propagators
of the $\psi$ and $\sigma$ fields where the form of the latter involves a
hypergeometric function of the momentum and mass, \cite{1}. (This
non-fundamental structure is consistent with its fermion bound state nature.)

As we have already mentioned in the self consistency approach of \cite{2} it
is indeed possible to go beyond the leading order by examining the structure
of the field theory in the neighbourhood of its $d$-dimensional critical
point defined as the non-trivial zero of the $\beta$-function. There the theory
possesses a conformal symmetry and the Green's functions of the fields obey a
simple power law behaviour and there is no mass in the problem. By solving
for the critical exponents of the power law one can deduce information on the
renormalization group functions through a critical point analysis of the
renormalization group equation which takes a solvable and simplified form at
criticality, \cite{5}. For this paper, taking $1/N$ as the coupling constant
allows us to compute the exponent order by order in $1/N$ by solving the
appropriate Dyson equations truncated to the order we are interested in.

We will now review the $O(1/N)$ derivation of $2\lambda$ $=$ $-$
$\beta^\prime(g_c)$ in the self consistency formalism to illustrate the
general method and to define the notation of the problem. As the $\sigma$ and
$\pi$ fields are dynamical, we take, \cite{6,15},
\begin{eqnarray}
\psi(x) & \sim & \frac{A{x \!\!\! /}}{(x^2)^\alpha}
[ 1 + (x^2)^\lambda A^\prime] \\
\sigma(x) & \sim & \frac{B}{(x^2)^\beta} [ 1 + (x^2)^\lambda B^\prime] \\
\pi(x) & \sim & \frac{C}{(x^2)^\gamma}[1+ (x^2)^\lambda C^\prime]
\end{eqnarray}
as the asymptotic scaling forms of the propagators of the fields in
coordinate space in the critical region as $x$ $\rightarrow$ $0$. The
quantities $A$, $B$ and $C$ and $A^\prime$, $B^\prime$ and $C^\prime$ are the
respective $x$-independent amplitudes, whilst $\alpha$, $\beta$ and
$\gamma$ are the full dimensions of the respective fields. We choose to
compute in coordinate space in order to implement various integration
techniques which are easier to apply there but note that one can access the
momentum space formulation easily via the Fourier transform
\begin{equation}
\frac{1}{(x^2)^\alpha} ~=~ \frac{a(\alpha)}{2^{2\alpha}\pi^\mu} \int_k
\frac{e^{ikx}}{(k^2)^{\mu-\alpha}}
\end{equation}
valid for all $\alpha$ where $a(\alpha)$ $=$
$\Gamma(\mu-\alpha)/\Gamma(\alpha)$ and we set the spacetime dimension $d$ to
be $d$ $=$ $2\mu$ for convenience. One can deduce the canonical dimensions
of $\alpha$, $\beta$ and $\gamma$ from a dimensional analysis of the action
but quantum fluctuations through, say, radiative corrections will always
adjust these. To allow for this one introduces the anomalous dimensions in
the definitions of the exponents via
\begin{equation}
\alpha ~=~ \mu + {\mbox{\small{$\frac{1}{2}$}}} \eta ~~,~~ \beta ~=~ 1 - \eta
- \chi_\sigma ~~,~~ \gamma ~=~ 1 - \eta - \chi_\pi
\end{equation}
where $\eta$ is the fermion anomalous dimension and $\chi_\sigma$ and
$\chi_\pi$ are the anomalous dimensions of the $3$-vertices of (2.1)

In \cite{6,15} we considered only the leading order terms of the asymptotic
scaling forms in the skeleton Dyson equations which allows one to deduce the
exponent $\eta$ at $O(1/N^2)$. To determine $\beta$-function corrections one
has to include the higher order corrections as we have in (2.2)-(2.4) which
is similar to previous analyses in other models, \cite{3,6}. Then to deduce
the exponents one writes down a set of equations which represent the Dyson
equations in the critical region and solves them self consistently. We have
illustrated the Dyson equations we need to consider for $\lambda_1$, where
$\lambda$ $=$ $\mu$ $-$ $1$ $+$ $\sum_{i=1}^\infty \lambda_i/N^i$, in fig. 1.
Ordinarily to compute, say, $\eta_1$ one truncates the equations at one loop,
\cite{6}. However, as was noted in \cite{17} for field theories where the basic
field is fermionic one has to include the two loop graphs of the $\sigma$ and
$\pi$ $2$-point functions where each has an $(x^2)^\lambda$ insertion from
(2.3)-(2.4) on the $\sigma$ or $\pi$ internal propagator. The reason for
their inclusion is due to a reordering of the powers of $N$ in the master
equation whose solution yields $\lambda_1$. The quantities $\psi^{-1}$,
$\sigma^{-1}$ and $\pi^{-1}$ in fig. 1 are the $2$-point functions and their
asymptotic scaling forms can be deduced from (2.2)-(2.4) by inverting each
expression in momentum space via (2.5) and they take the $x$-space forms,
\cite{6},
\begin{eqnarray}
\psi^{-1}(x) &\sim& \frac{r(\alpha-1){x \!\!\! /}}{A(x^2)^{2\mu-\alpha+1}}
[ 1 - A^\prime s(\alpha-1) (x^2)^\lambda ] \\
\sigma^{-1}(x) &\sim& \frac{p(\beta)}{B(x^2)^{2\mu-\beta}}
[ 1 - B^\prime q(\beta) (x^2)^\lambda ] \\
\pi^{-1}(x) &\sim& \frac{p(\gamma)}{C(x^2)^{2\mu-\beta}}
[ 1 - C^\prime q(\gamma) (x^2)^\lambda ]
\end{eqnarray}
where we define
\begin{eqnarray}
p(\alpha) &=& \frac{a(\alpha-\mu)}{\pi^{2\mu}a(\alpha)} ~~~,~~~
q(\alpha) ~=~ \frac{a(\alpha-\mu+\lambda) a(\alpha-\lambda)}
{a(\alpha-\mu) a(\alpha)} \nonumber \\
r(\alpha) &=& \frac{\alpha p(\alpha)}{(\mu-\alpha)} ~~~,~~~
s(\alpha) ~=~ \frac{\alpha(\alpha-\mu)q(\alpha)}{(\alpha-\lambda)
(\alpha-\mu+\lambda)}
\end{eqnarray}
If we formally denote the values of the $2$-loop integrals by $\Pi_{1A,B,C}$
and $\Lambda_{1A,B,C}$, where the subscript $A$, $B$ or $C$ denotes an
insertion on a $\psi$, $\sigma$ or $\pi$ line respectively, then with
(2.2)-(2.4) the Dyson equations of fig. 1 become
\begin{eqnarray}
0 &=& r(\alpha-1) [1-A^\prime s(\alpha-1)(x^2)^\lambda]
+ z[1+(A^\prime + B^\prime)(x^2)^\lambda] \nonumber \\
&+& y[1+(A^\prime + C^\prime)(x^2)^\lambda] \\
0 &=& \frac{p(\beta)}{N} [1-B^\prime q(\beta)(x^2)^\lambda]
+ 2z[1+2A^\prime (x^2)^\lambda] \\
&-& z^2 [ \Pi_1 + (x^2)^\lambda(A^\prime\Pi_{1A} + B^\prime\Pi_{1B})]
+ zy [ \Pi_2 + (x^2)^\lambda(A^\prime\Pi_{2A} + C^\prime\Pi_{2C})] \nonumber \\
0 &=& \frac{p(\gamma)}{N} [1-C^\prime q(\gamma)(x^2)^\lambda]
+ 2y[1+2A^\prime (x^2)^\lambda] \\
&-& y^2 [\Lambda_1 + (x^2)^\lambda(A^\prime\Lambda_{1A} + C^\prime
\Lambda_{1C})] + zy [\Lambda_2 + (x^2)^\lambda(A^\prime\Lambda_{2A} + B^\prime
\Lambda_{2B})] \nonumber
\end{eqnarray}
where we have set $z$ $=$ $A^2B$ and $y$ $=$ $A^2C$. Comparing powers of
$x^2$ the equations decouple into a set which become the consistency equations
for $\eta_1$ and give, \cite{15},
\begin{equation}
\eta_1 ~=~ - \, \frac{2\Gamma(2\mu-1)}{\Gamma(\mu-1)\Gamma(1-\mu)\Gamma(\mu)
\Gamma(\mu+1)}
\end{equation}
and a set involving $A^\prime$, $B^\prime$ and $C^\prime$. The solution of the
latter yield $\lambda_1$ where the relevant consistency equation is formed by
setting the determinant of the matrix formed by treating $A^\prime$,
$B^\prime$ and $C^\prime$ as independent basis vectors, to zero. In this
matrix one can examine the $N$ dependence of each element and observe that
$r(\alpha-1)s(\alpha-1)$ $=$ $O(1)$ whilst $p(\beta)q(\beta)$ $=$ $O(1/N)$.
Since $y$ and $z$ are both $O(1/N)$ from the $\eta_1$ equation it is easy to
see the necessity of including the higher order graphs $\Pi_{1B,C}$ and
$\Lambda_{1B,C}$. Direct calculation, however, reveals that $\Pi_{1B}$ $=$
$\Pi_{1C}$ $=$ $\Lambda_{1B}$ $=$ $\Lambda_{1C}$ $\equiv$ $\Pi$, where the
basic integral, $\Pi$, has been computed explicitly in the $O(N)$ model,
\cite{6,17}.  It turns out that in manipulating the determinant through
elementary row and column transformations and using, \cite{15},
\begin{equation}
z_1 ~=~ y_1 ~=~ \frac{\mu \Gamma^2(\mu)\eta_1}{4\pi^{2\mu}}
\end{equation}
that one obtains
\begin{equation}
0 ~=~ \det \left(
\begin{array}{ccc}
s(\alpha-1) & 1 & 1 \\
2 & q(\beta) & z\Pi \\
0 & 0 & q(\gamma) - z \Pi \\
\end{array}
\right)
\end{equation}
where the $\Pi$-type contribution has cancelled in the central element through
the additional symmetry present, compared with \cite{6,15}. Consequently, the
consistency equation which results from (2.16),
\begin{equation}
0 ~=~ q(\beta) s(\alpha-1) ~-~ 2
\end{equation}
does not require the value of $\Pi$ and we deduce that
\begin{equation}
\lambda_1 ~=~ - \, (2\mu-1) \eta_1
\end{equation}
which is in agreement with \cite{16} and establishes the correctness of our
procedure for this model. This is a non-trivial statement as a second
consistency equation would appear to emerge from (2.16) by setting the lower
right element to zero. We discard this as it does not give a result consistent
with the exponent calculated in the conventional large $N$ approach. Moreover,
the appearance of such additional solutions is an artefact of the formalism and
has been observed in other models where again it was appropriate to disregard
it, \cite{18}. Indeed from the transformations of the determinant we have made
this potentially alternative solution does not take into account any
contribution from the $\psi$ consistency equation, which again justifies our
stance.

\sect{Master equation.}

To proceed beyond (2.18) and to compute the $O(1/N^2)$ correction requires
that one expands the quantities $s(\alpha-1)$, $q(\beta)$ and $q(\gamma)$ to
the subsequent order and also includes the higher order corrections to the
Dyson equations. For the former this requires the values of the vertex
anomalous dimensions and $\eta_2$. They have been calculated in \cite{15} by
considering the subsequent corrections to the Dyson equations which determine
the fermion anomalous dimension as
\begin{eqnarray}
\eta_2 &=& \eta^2_1 \left[ \Psi(\mu) \, + \, \frac{2}{(\mu-1)}
\, + \, \frac{1}{2\mu} \right] \\
\chi_{\sigma \, 1} &=& \chi_{\pi \, 1} ~=~ 0
\end{eqnarray}
where $\Psi(\mu)$ $=$ $\psi(2\mu-1)$ $-$ $\psi(1)$ $+$ $\psi(2-\mu)$ $-$
$\psi(\mu)$ and $\psi(x)$ is the logarithmic derivative of the
$\Gamma$-function. By considering the scaling behaviour of the $3$-vertices at
$O(1/N^2)$ one also finds that
\begin{equation}
\chi_{\sigma \, 2} ~=~ \chi_{\pi \, 2} ~=~
- \, \frac{\mu^2(4\mu^2-10\mu+7)\eta^2_1}{2(\mu-1)^3}
\end{equation}
using methods developed from the earlier work of \cite{19,20}.

The main problem in determining $\lambda_2$, however, is the inclusion of the
higher order corrections. For the fermion equations this is straightforward in
that one includes the two loop Feynman graphs of fig. 2, \cite{8}, and these
together with the two loop graphs of fig. 1 are all that is required to
deduce $\eta_2$ from the leading order ans\"{a}tze of (2.2)-(2.4) and
(2.7)-(2.9). It is worthwhile detailing the resulting Dyson equation for
$\psi$ explicitly since it will illustrate several important points which
arise in the $\sigma$ and $\pi$ equations and may be obscured there. We have
\begin{eqnarray}
0 &=& r(\alpha-1) [1-A^\prime s(\alpha-1)(x^2)^\lambda]
+ z(x^2)^{\chi_\sigma}[1+(A^\prime + B^\prime)(x^2)^\lambda]
\nonumber \\
&+& y(x^2)^{\chi_\pi}[1+(A^\prime + C^\prime)(x^2)^\lambda] \nonumber \\
&+& z^2 (x^2)^{2\chi_\sigma}[\Sigma_1 + (x^2)^\lambda(A^\prime
\Sigma_{1A} + B^\prime\Sigma_{1B})] \nonumber \\
&-& 2yz (x^2)^{\chi_\sigma+\chi_\pi}[\Sigma_2 + (x^2)^\lambda(A^\prime
\Sigma_{2A} + B^\prime\Sigma_{2B} + C^\prime\Sigma_{2C})] \nonumber \\
&+& y^2 (x^2)^{2\chi_\pi}[\Sigma_3 + (x^2)^\lambda(A^\prime
\Sigma_{3A} + C^\prime\Sigma_{3C})]
\end{eqnarray}
where the mixed two loop graphs have the same value and we have not cancelled
the powers of $x^2$. The values $\Sigma$ and $\Sigma_{A,B,C}$ correspond to the
values of the integrals representing the Feynman graphs without symmetry
factors which have been displayed explicitly. In analysing similar equations at
$O(1/N^2)$ in other models the analogous higher order two loop graphs are
infinite which can be deduced by a detailed computation. To handle such
divergences one shifts the exponents of the $\sigma$ and $\pi$ fields by an
infinitesimal amount $\Delta$ through $\chi_\sigma$ $\rightarrow$ $\chi_\sigma$
$+$ $\Delta$ and $\chi_\pi$ $\rightarrow$ $\chi_\pi$ $+$ $\Delta$ where
$\Delta$ plays the role of a regularizing parameter. The subsequent simple
poles in $\Delta$ which would occur in the regularized graphs are renormalized
by the vertex counterterm available from the one loop graphs. For the current
equation it turns out that upon adding the divergent graphs together the
infinity cancels due to the symmetry induced through the $\gamma^5$ interaction
and there is therefore no need to renormalize (3.4) to the order we are working
to, $O(1/N^2)$. Thus only the $\Delta$-finite parts of the $2$-loop graphs
remain in (3.4). As before we decouple the finite equation into that which has
already been used to deduce $\eta_2$ and
\begin{eqnarray}
0 &=& [ z+y-r(\alpha-1)s(\alpha-1) + z^2 \Sigma_{1A} - 2yz \Sigma_{2A}
+ y^2\Sigma_{3A}] A^\prime \nonumber \\
&+& [z + z^2\Sigma_{1B} - 2yz \Sigma_{2B}] B^\prime
+ [y + y^2\Sigma_{3C} - 2yz \Sigma_{3C}] C^\prime
\end{eqnarray}
where the powers of $x^2$ are absent at this order due to (3.2). The explicit
evaluation of the finite parts of $\Sigma_{iB}$ and $\Sigma_{iC}$, however,
have been determined in \cite{8} and it turns out that to the order we are
working to they are zero. Also if one compares the $N$-dependence of each term
of the coefficient of $A^\prime$ in (3.5) it is easy to see that $\Sigma_{iA}$
is $O(1/N^2)$ relative to $r(\alpha-1)s(\alpha-1)$ and therefore can be
neglected in the correction to the determinant of (2.16). This term would be
relevant only for $\lambda_3$.  Thus the part of (3.5) which we require for
$\lambda_2$ is simply
\begin{equation}
0 ~=~ [ z + y - r(\alpha-1) s(\alpha-1)] A^\prime + z B^\prime + y C^\prime
\end{equation}
where, of course, the $O(1/N^2)$ parts of $z$ and $y$ contribute in the last
two terms.

At leading order we had to include the higher order two loop graphs of fig.
1. As was noted in \cite{17} this is a feature of the formalism for models
where the fundamental field is fermionic, although in the final equation,
(2.17), for the chiral Gross Neveu model the explicit value was not needed.
By the same argument one has now to include the representative three and four
loop higher order graphs of fig. 3 in the $\sigma$ and $\pi$ equations which
we now discuss. We have given only the basic distinct topological structures
of the graphs which occur for $\sigma$ where the internal bosonic lines can
be either the $\sigma$ or $\pi$ fields. So, for example, there are $4$, $4$,
$4$, $8$ and $8$ respective graphs for each of the basic ones given in fig.
3 for the $\sigma$ equation. Of course, since $\mbox{tr}\,\gamma^5$ $=$ $0$
only half of the four loop graphs survive. Further, when one substitutes the
asymptotic scaling forms (2.3) and (2.4) one need only include those graphs
in the $\sigma$ consistency equation where there is one $(x^2)^\lambda$
insertion on an internal bosonic line, in a similar way to leading order,
\cite{8}. The full consistency equation for $\sigma$ with the graphs of
fig. 3 is quite a long expression and rather than give its complete form we
write down only the part relevant for $\lambda_2$ as
\begin{equation}
0 ~=~ 4zA^\prime - B^\prime \left[ \frac{p(\beta)q(\beta)}{N} + z^2 \Pi
+ \Pi_{B2} \right] + C^\prime [ yz \Pi - \Pi_{C2} ]
\end{equation}
where we have set
\begin{eqnarray}
\Pi_{B2} &=& 2(\Pi_{2B1} + \Pi_{2B2}) + \Pi_{3B} + \Pi_{4B} \nonumber \\
&-& 2(2\Pi_{5B1} + \Pi_{5B2}) - 4(\Pi_{6B1} + 2\Pi_{6B2}) \\
\Pi_{C2} &=& 2(\Pi_{2C1} + \Pi_{2C2}) + \Pi_{3C} + \Pi_{4C} \nonumber \\
&-& 2(2\Pi_{5C1} + \Pi_{5C2}) - 4(\Pi_{6C1} + 2\Pi_{6C2})
\end{eqnarray}
and our notation is partially defined in fig. 3. When there is an additional
subscript $1$ or $2$ in (3.8) and (3.9) this corresponds to the two distinct
ways of including an insertion on the internal bosonic line and is consistent
with definitions given in \cite{8}. Using the properties of the
$\gamma$-matrices and being careful in manipulating factors of $i$ from the
vertex of (2.1) one finds that the contributions from $\Pi_{2B1}$, $\Pi_{3B}$,
$\Pi_{2C1}$ and $\Pi_{3C}$ style graphs each sum to zero through use of (2.15).
This is consistent with the fact that one does not require a regularization at
this order in $1/N$ as the divergent contributions cancel. Subsequently this
implies $\chi_{\sigma \, 1}$ $=$ $0$ which is in agreement with calculations
of other Dyson equations. Further, with (2.15) it is possible to simplify
(3.8) and (3.9) to
\begin{eqnarray}
\Pi_{B2} &=& 4z^3[\Pi^{\mbox{o}}_{2B2} + \Pi^{\mbox{o}}_{4B}
- z_1(2\Pi^{\mbox{o}}_{5B1} + \Pi^{\mbox{o}}_{5B2}
+ 2\Pi^{\mbox{o}}_{6B1})] \\
\Pi_{C2} &=& 4z^3[- \Pi^{\mbox{o}}_{2B2} + \Pi^{\mbox{o}}_{4B}
- z_1(2\Pi^{\mbox{o}}_{5B1} + \Pi^{\mbox{o}}_{5B2}
- 2\Pi^{\mbox{o}}_{6B1})]
\end{eqnarray}
using (2.15) where the superscript ${}^{\mbox{o}}$ denotes the value of the
fundamental graph, with the appropriate insertion, given in fig. 3. Whilst
these have all been computed explicitly in the $O(N)$ case, \cite{8}, we defer
to later the direct substitution into $\Pi_{B2}$ and $\Pi_{C2}$.

A similar analysis of the $\pi$ Dyson equation yields the third of our
consistency equations as
\begin{equation}
0 ~=~ 4y A^\prime + B^\prime[ yz \Pi - \Pi_{C2}] - C^\prime
\left[ \frac{p(\gamma)q(\gamma)}{N} + y^2 \Pi + \Pi_{B2} \right]
\end{equation}
where the same combinations $\Pi_{B2}$ and $\Pi_{C2}$ appear as in (3.7). We
can now manipulate these formal corrections to (2.11)-(2.13) to deduce the
master equation for $\lambda_2$. First, though we record that from the
$\eta_2$ consistency equation, \cite{15},
\begin{equation}
z_2 ~=~ y_2 ~=~ \frac{\mu \Gamma^2(\mu)\eta^2_1}{4\pi^{2\mu}}
\left[ \Psi(\mu) + \frac{2}{(\mu-1)} \right]
\end{equation}
and following the same transformations of the determinant as we made at leading
order with these corrections we are left with the determinant of the $2$
$\times$ $2$ submatrix
\begin{equation}
0 ~=~ \det \left(
\begin{array}{cc}
2z-r(\alpha-1)s(\alpha-1) & 2z \\
4z & - {\mbox{\small{$\frac{1}{N}$}}} p(\beta)q(\beta) - (\Pi_{B2} + \Pi_{C2})
\\
\end{array}
\right)
\end{equation}
Again the two loop graphs have cancelled though it is important to note that
they would in fact have contained $O(1/N^2)$ information since the integral
depends on the exponents $\eta_1$ and $\lambda_1$ and they would have
contributed to $\lambda_2$. From (3.14) we deduce that $\lambda_2$ will emerge
from the solution of
\begin{equation}
8 z^2 ~=~ [r(\alpha-1) s(\alpha-1) - 2z] \left[ \frac{p(\beta)q(\beta)}{N}
+ \Pi_{B2} + \Pi_{C2} \right]
\end{equation}
Again we ignore the solution that would come from the lower right element of
the full $3$ $\times$ $3$ for the reasons stated earlier. Moreover, we note
that the higher order graph contributions appear in a particular combination
$\Pi_{B2}$ $+$ $\Pi_{C2}$ and from (3.10) and (3.11) this simplifies to
\begin{equation}
\Pi_{B2} + \Pi_{C2} ~=~ 8z^3 [\Pi^{\mbox{o}}_{4B} - z_1(2\Pi^{\mbox{o}}_{5B1}
+ \Pi^{\mbox{o}}_{5B2})]
\end{equation}
Thus we need only consider the contributions from three of the eight basic
topologies and insertions of fig. 3 which represents a significant
simplification. We note that the graph $\Pi^0_{5B1}$ corresponds to an
$(x^2)^\lambda$ insertion on the top or bottom internal bosonic line.

\sect{Discussion.}

All that remains now is the explicit evaluation of (3.15) at $O(1/N^2)$. The
values of the higher order graphs have been computed in \cite{8} for the Gross
Neveu model with a discrete chiral symmetry. This involved extensive use of the
technique known as uniqueness, \cite{4}, which is applicable to calculating
massless Feynman diagrams, and is a method of integration which is based on the
conformal properties of the integral in $d$-dimensions. Our manipulations for
(2.1) have been such that the basic graphs of (3.16) are equivalent to the
values of the graphs of the $O(N)$ case and therefore we record that, \cite{8},
\begin{equation}
\Pi^{\mbox{o}}_{4B} ~=~ \frac{\pi^{4\mu}}{(\mu-1)^2\Gamma^4(\mu)} \left[
3\Theta
+ \frac{1}{(\mu-1)^2} \right]
\end{equation}
\begin{eqnarray}
\Pi^{\mbox{o}}_{5B1} &=& \frac{\pi^{6\mu}(2\mu-3)a(2\mu-2)}
{2(\mu-1)^5(\mu-2)\Gamma^3(\mu)} \left[ 6 \Theta - \Phi - \Psi^2
+ \frac{5}{2(\mu-1)^2} \right. \nonumber \\
&+& \left. \frac{\Psi}{(2\mu-3)(\mu-1)} + \frac{2(\mu-2)\Psi}{(\mu-1)}
+ \frac{1}{(2\mu-3)(\mu-1)} \right. \nonumber \\
&-& \left. \frac{8}{(2\mu-3)} + \frac{(\mu-2)}{(\mu-1)^2} \right]
\end{eqnarray}
\begin{eqnarray}
\Pi^{\mbox{o}}_{5B2} &=& \frac{\pi^{6\mu}a(2\mu-2)}{(\mu-1)^5 \Gamma^3(\mu)}
\left[ \frac{(2\mu-3)}{(\mu-2)} \left( \Phi - \Psi^2 - \frac{1}{2(\mu-1)^2}
\right) \right. \nonumber \\
&-& \left. \frac{(3\mu-4)\Psi}{(\mu-1)(\mu-2)^2} + \frac{1}{(\mu-2)^2}
\right] + \frac{2\pi^{6\mu}a(2\mu-2)}{(\mu-1)^6(\mu-2)^2}
\end{eqnarray}
where $\Theta(\mu)$ $=$ $\psi^\prime(\mu)$ $-$ $\psi^\prime(1)$ and
$\Phi(\mu)$ $=$ $\psi^\prime(2\mu-1)$ $-$ $\psi^\prime(2-\mu)$ $-$
$\psi^\prime(\mu)$ $+$ $\psi^\prime(1)$. Thus substituting into (3.16) we
have
\begin{eqnarray}
\Pi_{B2} + \Pi_{C2} &=& \frac{8\pi^{4\mu}z_1^3}{(\mu-1)^2\Gamma^4(\mu)}
\left[ \frac{2(2\mu-3)}{(\mu-2)} (\Phi + \Psi^2)
- \frac{3(3\mu-4)\Theta}{(\mu-2)} \right. \nonumber \\
&+& \left. \frac{8(\mu-1)^2}{\mu(\mu-2)^2\eta_1}
- \frac{2(2\mu-3)(\mu^2-4\mu+5)\Psi}{(\mu-1)(\mu-2)^2}
+ \frac{1}{(\mu-1)^2} \right. \nonumber \\
&-& \left. \frac{2(2\mu-3)}{(\mu-2)(\mu-1)^2} + \frac{6}{(\mu-2)}
+ \frac{1}{(\mu-2)^2} \right]
\end{eqnarray}
A little bit of algebra subsequently leads us to the main result of the
paper,
\begin{eqnarray}
\lambda_2 &=& \eta^2_1 \left[ \frac{3\mu^2(3\mu-4)\Theta}{2(\mu-1)(\mu-2)}
- \frac{\mu^2(2\mu-3)(\Phi + \Psi^2)}{(\mu-1)(\mu-2)}
+ \frac{4\mu(\mu-1)}{(\mu-2)^2\eta_1} \right. \nonumber \\
&+& \left. \frac{2}{(\mu-1)} + \frac{1}{2\mu}
- \frac{\mu^2(2\mu-3)(2\mu^3-9\mu^2+16\mu-11)}{2(\mu-1)^3(\mu-2)^2} \right.
\nonumber \\
&+& \left. 2\mu(2\mu-3) + \Psi \left( \frac{\mu^2(2\mu-3)(\mu^2-4\mu+5)}
{(\mu-1)^2 (\mu-2)^2} - 4\mu^2 +1 \right) \right]
\end{eqnarray}
in arbitrary dimensions.

As the three dimensional chiral Gross Neveu model has also been the subject of
study recently, \cite{22,16,21}, we can set $\mu$ $=$
${\mbox{\small{$\frac{3}{2}$}}}$ in (4.5) to deduce a new result for this case
as
\begin{equation}
\lambda ~=~ \frac{1}{2} - \frac{16}{3\pi^2N} + \frac{16[472+27\pi^2]}{27\pi^4}
\end{equation}
This will be important for deducing numerical estimates for this exponent
for relatively small values of $N$ since several orders are now known. For
example, in \cite{8} an estimate was given for the $O(8)$ Gross Neveu model
which agreed with a recent Monte Carlo result, \cite{22}.

We conclude with various remarks. First, the chiral Gross Neveu model has now
been solved completely at $O(1/N^2)$ and it is important to recognise that this
is one order further than had previously been possible. By this we mean that
two independent critical exponents $\lambda$ and $\eta$ or $\eta$ $+$
$\chi_\sigma$ have been determined to this order. The remaining thermodynamic
exponents can be deduced through the hyperscaling laws which are valid for
renormalizable quantum field theories and which have been checked in \cite{16}.
The quantum structure of the model is appreciably simpler than for the $O(N)$
Gross Neveu model in the sense that fewer topologies of Feynman diagrams needed
to be considered and this is primarily due to the extra chiral symmetry present
in (2.1). Indeed a clear indication of symmetry within the self consistency
formalism is the vanishing of the anomalous dimensions of one of the fields or
vertices, such as $\chi_{\sigma \, 1}$ $=$ $\chi_{\pi \, 1}$ $=$ $0$ here, in
all dimensions. A similar phenomena occurs in the supersymmetric $O(N)$
$\sigma$ model where the anomalous dimension of the analogous supermultiplet of
auxiliary and Lagrange multiplier fields vanishes at $O(1/N)$ as a consequence
of the unbroken supersymmetry present in the model, \cite{23}.

\vspace{1.5cm}
\noindent
{\Large {\bf Figure Captions.}}
\begin{description}
\item{Fig. 1.} Leading order Dyson equations to determine $\lambda_1$.
\item{Fig. 2.} Higher order graphs for $\psi$ Dyson equation.
\item{Fig. 3.} Higher order graphs for $\sigma$ and $\pi$ Dyson equations.
\end{description}
\end{document}